\documentclass[preprint,12pt,authoryear]{elsarticle}
\usepackage{amssymb}
\usepackage{lineno}
\usepackage{booktabs}
\usepackage{bm}
\usepackage{multirow}
\usepackage[table,xcdraw]{xcolor}
\usepackage{xr-hyper}
\usepackage{hyperref}
\usepackage{gensymb}
\usepackage{xcolor}
\usepackage{soul}
\usepackage[normalem]{ulem}
\usepackage{booktabs}
\usepackage{multirow}

\journal{Atmospheric Environment}

\begin{document}
\begin{frontmatter}

  \title{Calibrating the CAMS European  multi-model air quality forecasts for regional air pollution monitoring}

\author[DICCA]{Gabriele Casciaro} \ead{gabriele.casciaro@edu.unige.it} 

\author[DICCA,INFN]{Mattia Cavaiola} \ead{mattia.cavaiola@edu.unige.it}

\author[DICCA,INFN]{Andrea Mazzino\corref{cor}} \ead{andrea.mazzino@unige.it}

\affiliation[DICCA]{organization={DICCA, Department of Civil, Chemical and Environmental Engineering. University of Genoa},
            addressline={Via Montallegro 1}, 
            city={Genoa},
            postcode={16145}, 
            state={Genoa},
            country={Italy}}
\affiliation[INFN]{
            organization={INFN, Nazional Institute for Nuclear Physics, Genoa Section},            addressline={Via Dodecaneso 33}, 
            city={Genoa},
            postcode={16146}, 
            state={Genoa},
            country={Italy}}

\cortext[cor]{Corresponding author}

\begin{abstract}
The CAMS air quality multi-model forecasts have been assessed and calibrated for  PM$_{10}$, PM$_{2.5}$, O$_{3}$, NO$_{2}$, and CO against observations collected by
the Regional Monitoring Network of the Liguria region (northwestern Italy) in the years 2019 and 2020.
The calibration strategy used in the present work has its roots in the well-established
Ensemble Model Output Statistics (EMOS) through which a raw ensemble forecast can be accurately transformed into a predictive probability density function, with a simultaneous  correction of biases and dispersion errors. The strategy also provides a calibrated forecast of model uncertainties.
As a result of our analysis, the key role of pollutant real-time observations to be ingested in the calibration strategy clearly emerge especially in the shorter look-ahead forecast hours.
Our dynamic calibration strategy turns out to be superior with respect to its analogous where real-time data are not taken into account.
The best calibration strategy we have identified makes the CAMS multi-model forecast system
more reliable than other raw air quality models running at higher spatial resolution which exploit
more detailed information from inventory emission.
We expect positive impacts of our research for identifying and set up reliable and economic
 air pollution  early warning systems.
\end{abstract}




\begin{highlights}

\item A dynamic calibration strategy brings out the intrinsic strength of CAMS multi-model forecasts
\item The CAMS calibrated forecasts neatly outperforms persistence
\item The CAMS calibrated forecasts has larger skills than raw higher-resolution strategies 
\item The CAMS calibrated system also provides calibrated forecasts of model uncertainties
\item Our outcomes impact the management of air pollution  early warning systems

\end{highlights}


\begin{keyword}

Ensemble forecasts \sep Air quality monitoring \sep Ensemble model output statistics (EMOS) \sep Static vs. dynamic calibrations  \sep  Air pollution  early warning systems

\end{keyword}

\end{frontmatter}


\section{Introduction}
\label{Introduction}

Among all environmental problems, air pollution is surely one of the most
serious causing, each year,  at least 7 million premature deaths.
There are statistical evidences of positive correlations between deaths caused by air pollution and the ground concentration of fine particulate matter, PM,
\citep{yin2017particulate}. Studies have also shown that PM, ozone, nitrogen dioxide, and sulphur dioxide degrade ambient air quality and cause severe health problems in humans \citep{yin2017particulate}. In way of example, 
increased ground-level ozone is responsible for metabolic effects on humans leading to glucose intolerance and hyperlipidemia \citep{shore2019metabolic}. \\
Following the \cite{world2016bank}, these dramatic consequences that affect people's lives
also have huge implications for the world economies. The damage to the local economies are estimated \citep{world2016bank} in US \$ 5 trillion in welfare losses and US \$ 255 billion in lost labor income.\\
Ecosystems are also impacted by air pollution, particularly sulphur and nitrogen emissions, and ground-level ozone as it affects their ability to function and grow  \citep{wang2021satellite}. Emissions of both sulphur dioxide and nitrogen oxides deposit in water, on vegetation and on soils as `acid rain', thereby increasing their acidity with adverse effects on flora and fauna \citep{vries2014impacts}.
Increased ground-level ozone also causes damage to cell membranes on plants inhibiting key processes required for their growth and development \citep{jae2020onsite}.\\
Ozone and PM also interact with radiation thus forcing climate change \citep{fiore2015air,denjean2016size}. PM warms by absorbing sunlight (e.g., black carbon) or cools by scattering sunlight (e.g., sulfates) and interacts with clouds; these radioactive and microphysical interactions can induce changes in precipitation and regional circulation patterns.\\
All the above considerations clearly tell us that air quality is far from being a `local' problem: it affects human life in a myriad of different ways and the same is
for the whole Earth environment, a fact requiring effective, global, pollution control/prevention measures. These latter are nowadays possible thanks to the concomitant advent of supercomputers and of accurate air pollution forecasting strategies.
Observations, both from satellites and from the ground, can provide a snapshot of the air quality, but have no real predictive capability, at least on time horizons of the order of a few days.  Combining state-of-the-art numerical models of the atmosphere with satellite and ground-station observations to provide daily forecasts of the composition of the air worldwide, has become a key task to set up  strategies of pollution control and risk prevention.
 It is in this framework that the Air Quality thematic area  of CAMS, the Copernicus Atmosphere Monitoring Service, monitors and forecasts European air quality long-range transport of pollutants \citep{CAMS}.
 The combination of millions of daily observations and the predictive power of Numerical Prediction Models (NPM) is a real strength of CAMS. But this is not the only one. CAMS air quality forecasts indeed merge the power of deterministic forecasts together with the power of statistical forecasting. CAMS forecasts are indeed based on an ensemble of 9 state-of-the-art numerical air quality models developed in Europe: CHIMERE from INERIS (France), EMEP from MET Norway (Norway), EURAD-IM from J\"ulich IEK (Germany), LOTOS-EUROS from KNMI and TNO (Netherlands), MATCH from SMHI (Sweden), MOCAGE from Meteo-France (France), SILAM from FMI (Finland), DEHM from Aarhus University (Denmark), and GEM-AQ from IEP-NRI (Poland).
End-users have in this way information about prediction uncertainty together with a single forecast for different pollutants (corresponding, e.g., to the median of the empirical distribution composed by the 9 members) for each forecast time horizon.\\
There is however a  price to pay for obtaining worldwide air quality forecasts. The necessarily coarse spatial  resolution (of about 0.1$^o$ for CAMS)
indeed implies that
small-scale features, including small-scale orography variations, small-scale meteorological conditions and related dispersion regimes, local pollutant emissions not captured by large-scale emission inventories, and many others, are not explicitly accounted for in large-scale air-quality forecast models. Using `raw' CAMS forecasts to predict pollutant ground concentrations at local scale may thus results in large forecast errors.  The mean and variance of the ensemble is expected to correlate with observations and the actual model uncertainty, respectively, but a tendency to underestimate and to be underdispersive with respect to the true observations and uncertainty of the model is also expected. This is a reasonable expectation in analogy with what happens for the 50-member-based meteorological Ensemble Prediction System (EPS) of the European Centre for Medium-Range Weather Forecasts (ECMWF)
\citep{molteni1996ecmwf, montani2019performance}.
The lesson we learned from the ECMWF EPS is that, used as is to predict, e.g., winds in regions of complex orography, has a limited skill. On the contrary, accurate predictions can be obtained via suitable statistical post processing methods of the raw predictions which use past observed data to train the calibration algorithm \citep{casciaro2022novel}.
\begin{figure}[h!]
\includegraphics[width=\textwidth]{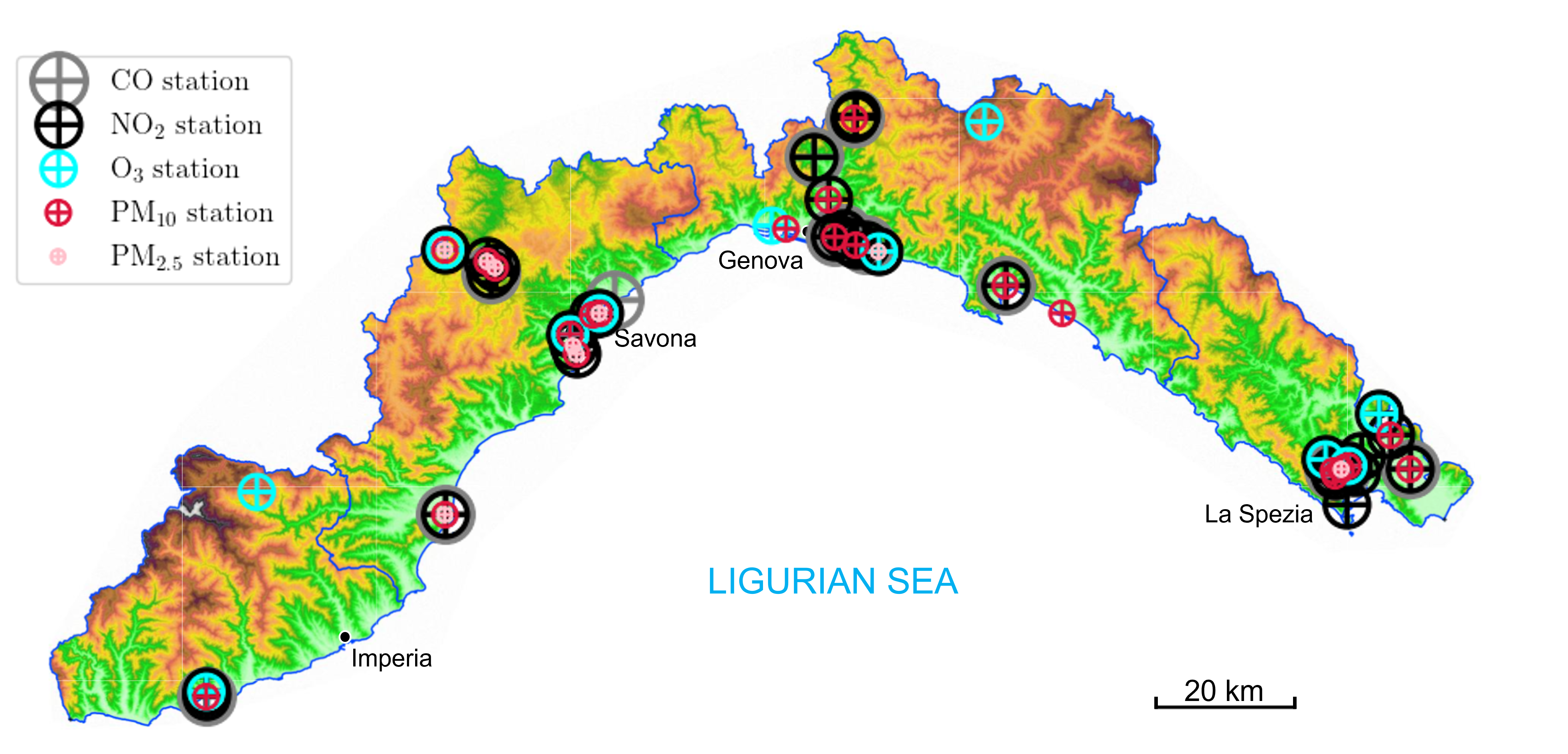}
\caption{The Liguria region (northwestern Italy) hosting the Regional Monitoring Network used here to calibrate/validate the CAMS air quality forecasts.}
\label{fig:liguria}
\end{figure}

In this study, we explore the possibility of using CAMS air quality forecasts to predict the 48-hour ahead evolution of PM$_{10}$, PM$_{2.5}$ (daily averages), O$_{3}$, NO$_{2}$, and CO (hourly averages) in a network of stations located in Liguria (northwestern Italy).  
The choice of the Ligurian region is motivated from the fact that Liguria presents a large variety in its orographic shape (see Fig.\ \ref{fig:liguria}), ranging from the alpine region (Alpi Marittime), with elevations up to 2000 meters, to flat areas as, e.g., in most zones close to the sea (Ligurian Sea). Moreover, the mutual interaction between land and sea circulations makes wind predictions, and predictions of weather variables in general, a very hard problem. Because weather conditions
heavily constrain air quality forecasts, the same difficulties, probably even more severely, also affect pollutant forecasts.
As we will see, for the first 24-hour-ahead forecast horizon, the added value of the raw CAMS forecasts with respect to persistence is very limited (if there is one).\\ 
Here, we propose a statistical calibration of CAMS ensemble
based on the use of past pollutant measurements collected by the stations
of the Ligurian monitoring network
(see Fig.\ \ref{fig:liguria}) in the years 2019 and 2020.
The calibration strategy generalizes to air quality the one recently
proposed by \cite{casciaro2022novel} to calibrate the
raw EPS forecasts  
for the wind speed and wind power in complex orography regions for the green energy market.\\
As we will see, the calibration strategy remarkably increases the forecast skill, thus making CAMS global forecasts appealing also for monitoring activities at local scales.\\
Although our strategy has been applied to Liguria as an example, 
it can be trivially extended to any other area in the world. 
For our calibration strategy to be used operatively, the observed pollutant concentration  on the site of interest must be available in real/quasi-real time, other than as a record of pairs of past observations/past forecasts.\\
The paper is organized as follow.
In Sec.\ \ref{sec:CAMS} a short introduction to the CAMS multi-model ensemble system is presented;
Sec.\ \ref{sec:EPS} presents the assessment of raw forecasts of the CAMS multi-model system
against observations collected by the regional monitoring service of the Liguria region;
Sec.\ \ref{sec:dynamic} introduces our calibration strategies, static and dynamic, while our results are shown in Sec.\ \ref{Sec:results}. Concluding remarks and perspectives are drawn in the final section.

\section{The CAMS air quality forecasts in a nutshell}
\label{sec:CAMS}
The CAMS air quality models provide daily forecasts of the main atmospheric pollutants concentrations on daily basis and with a look-ahead forecast horizon of 4 days with a  spatial resolution of 0.1 degrees and 7 vertical levels from the Earth surface up to 5000 m. CAMS ensemble prediction system is composed by 9 distinguishable air quality models, expression of state-of-the-art knowledges in the field of
multi-scale chemistry-transport models for atmospheric composition. 
From the outputs of the 9 individual models, the ensemble  median can be calculated  yielding forecast performances higher than that of single models.
The  strength of CAMS ensemble lies in the fact that it can provide
an estimate of the forecast uncertainty via the spread between the
9 models.\\
The 9 models are distinguishable not only for the different details of the physics/chemistry they describe but also for the fact that they rely on their own data assimilation system. This latter consists of 
daily analyses of pollutants near the surface by merging 1-day old observations with model forecasts.\\
The meteorological fields needed to carry out the transport stage of pollutants come from the 00:00 UTC operational ECMWF IFS forecast from the day before, the boundary conditions for chemical species come from the CAMS IFS-TM5 global production, while the emissions come from CAMS emission (for anthropic emissions over Europe and for biomass burning).\\
CAMS forecasts are available for download from the CAMS Atmosphere Data Store (ADS). The production time is 05:50 UTC (for 0-24 h forecasts) and 05:55 UTC   (for 24-48 h forecasts) with data availability guaranteed by 08:00 UTC, thus being usable for operative needs.

Details on the multi-model forecast system are provided in \cite{CAMS}.

\section{Assessing the raw CAMS air quality forecasts}
\label{sec:EPS}
Let us start our analysis by assessing the raw CAMS air quality forecasts against
measured data collected and validated by the monitoring service of the Liguria region (see Fig.\ \ref{fig:liguria}).
The following pollutants have been considered in the present study:
PM$_{10}$ (daily averages), PM$_{2.5}$ (daily averages), O$_{3}$ (hourly averages), NO$_{2}$ (hourly averages), and CO (hourly averages). Only 7 CAMS multi-model ensemble members have been considered, GEM-AQ and DEHM becoming available only from the last three months of 2019.
\begin{table}[h!]
\caption{Details on the observation stations grouped together by pollutant type.}
\centering
\begin{tabular}{|c|c|c|c|}
\hline
Pollutant & \begin{tabular}[c]{@{}c@{}}Number of\\  stations\end{tabular} & \begin{tabular}[c]{@{}c@{}}Data coverage\\  (\%) 2019\end{tabular} & \begin{tabular}[c]{@{}c@{}}Data coverage\\ (\%) 2020\end{tabular} \\ \hline
CO        & 13                                                            & 56                                                                 & 73                                                                \\ \hline
NO$_2$       & 34                                                            & 74                                                                 & 63                                                                \\ \hline
O$_3$        & 11                                                            & 77                                                                 & 71                                                                \\ \hline
PM$_{10}$      & 22                                                            & 83                                                                 & 80                                                                \\ \hline
PM$_{2.5}$     & 9                                                             & 84                                                                 & 80                                                                \\ \hline
\end{tabular}
\label{tab:stations}
\end{table}

\begin{figure}[h!]
\includegraphics[width=\textwidth]{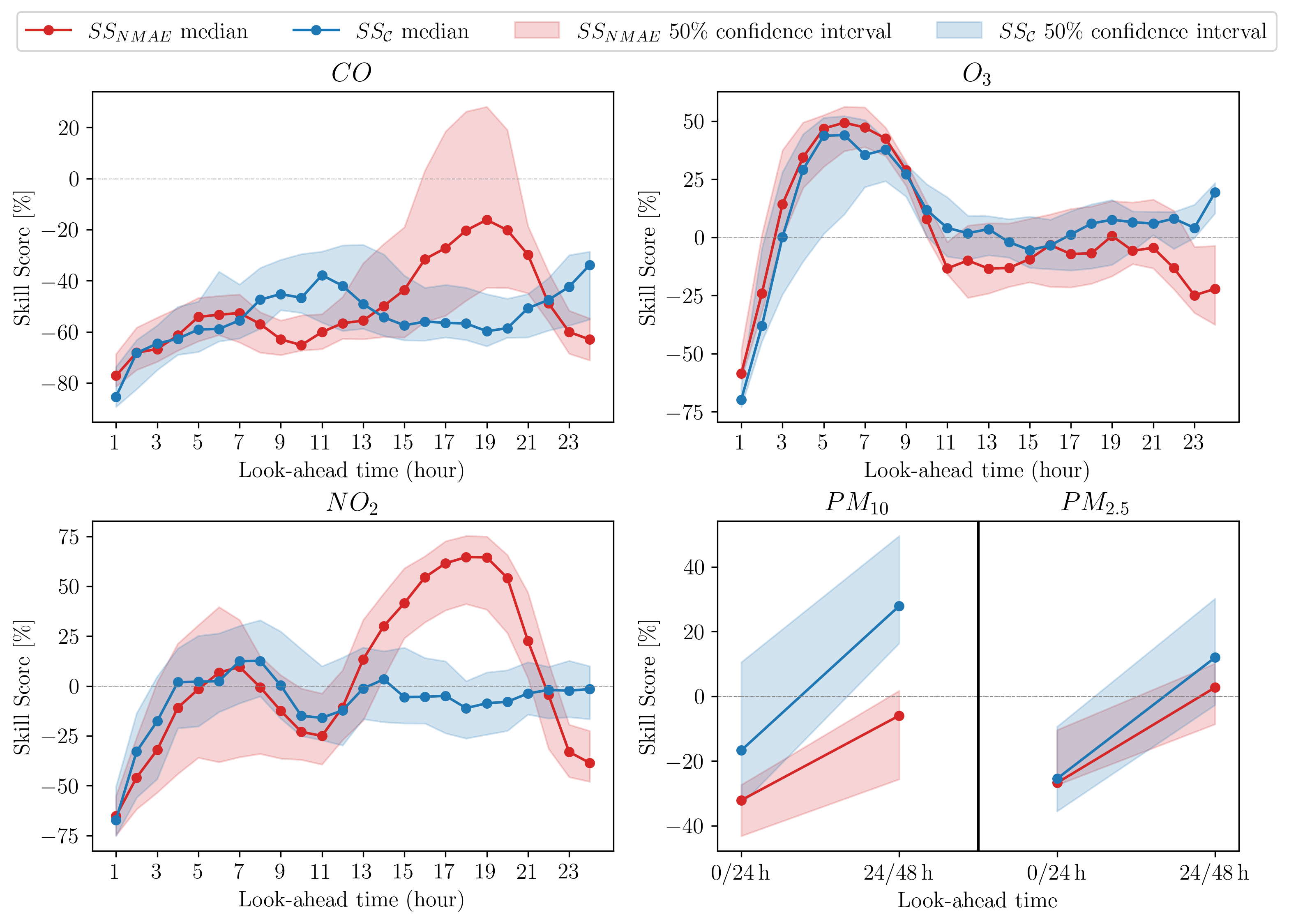}
\centering
\caption{
  Mean over  all stations  of the skill  score  for  both  the NMAE index and the correlation coefficient, ${\cal C}$, as a function of the look-ahead time, for the mean of the raw CAMS ensemble forecasts
The  reference  forecast to calculate the skill score is based on persistence built from observations at 9 UTC.  
Continuous lines: median of the skill scores computed by considering all stations. Shaded areas represent the 50\% confidence interval of the skill score, accounting for the variability of the latter
from station to station.}
\label{fig:raw_vs_pers00}
\end{figure}

In Tab.\ \ref{tab:stations} we report the number of ground stations for  each of the
measured pollutants  together with the total coverage for each of the two years considered, 2019 and 2020 (i.e.\ the number of validated observations divided by the maximum number of possible observations).
The assessment of raw CAMS forecasts is done in terms of standard
error indices, both to assess the accuracy of the mean (or median)  of the ensemble and  to quantify the accuracy of the whole probability distribution of the multi-model ensemble forecasts. The used indices are described in \ref{App:stat}. \\
Let us start by assessing the possible added value of CAMS forecasts against the most economic and simple forecast one can
do: the persistence based on observations at a given time. When dealing with model evaluation this step is strongly recommended.
Here we selected the hour 9 UTC to build the persistence-based forecasts.
The results are reported in Fig.\ \ref{fig:raw_vs_pers00} where the quality assessment of the CAMS
ensemble mean against persistence is evaluated in terms of the symmetrized skill score described in  \ref{App:stat}. For the sake of clarity, large positive values of the skill score
mean that the CAMS forecasts bring an added value with respect to persistence. The added value becomes more and more  important as the skill score approaches one. Negative values of the skill score mean that, on average,  persistence overcomes the raw CAMS forecast. 
For CO, O$_3$ and NO$_2$ the look ahead time along the abscissa is measured with respect to 9 UTC, the time used to define the persistence-based forecasts. An abscissa of 1 thus means the look-ahead time corresponding to 10 UTC. 
For PM$_{10}$ and PM$_{2.5}$ the look ahead time is with respect to the analysis time at 00 UTC.
As one can clearly see from the figure, the raw CAMS multi-model ensemble forecasts have a little added value with respect to persistence. The conclusion holds more or less for all pollutants here considered, more severely for CO and less dramatically  for the PMs in the forecast interval 24-48 h.
A complete panorama of the skill of the CAMS raw ensemble is presented in \ref{Confronto modelli raw}  where 
all single raw members are assessed for the year 2020 by averaging the single station indices, grouped together by pollutant type, over all stations.

The main conclusion we can draw from this first step of our analysis is that CAMS forecasts need to be calibrated using a training set of past observation-forecast pairs. Before to discuss this important issue, let us try to understand
how CAMS multi-model ensemble performs with respect to the model strategy, here referred to as S2020, used in \cite{sartini2020emission} having higher spatial resolution and accounting more detailed emission information from the regional inventory emissions. 

\begin{table}[h!]
\centering
\caption{A comparison between the hourly-averaged 24-h look ahead  raw forecasts from CAMS mean and from the high-resolution strategy reported in \cite{sartini2020emission}. Error indices are calculated in both cases against one year of observations of the Ligurian monitoring network.}
\resizebox{\textwidth}{!}{%
\begin{tabular}{|c|cccc|cccc|cccc|}
\hline
\multirow{3}{*}{\begin{tabular}[c]{@{}c@{}}Station\\ name\end{tabular}} & \multicolumn{4}{c|}{NO$_2$}                                                                        & \multicolumn{4}{c|}{O$_3$}                                                                        & \multicolumn{4}{c|}{PM$_{10}$}                                                                       \\ \cline{2-13} 
                                                                        & \multicolumn{2}{c|}{HH}                                  & \multicolumn{2}{c|}{NBI}             & \multicolumn{2}{c|}{HH}                                  & \multicolumn{2}{c|}{NBI}            & \multicolumn{2}{c|}{HH}                                  & \multicolumn{2}{c|}{NBI}             \\ \cline{2-13} 
                                                                        & \multicolumn{1}{c|}{Raw}  & \multicolumn{1}{c|}{S2020} & \multicolumn{1}{c|}{Raw}   & S2020 & \multicolumn{1}{c|}{Raw}  & \multicolumn{1}{c|}{S2020} & \multicolumn{1}{c|}{Raw}  & S2020 & \multicolumn{1}{c|}{Raw}  & \multicolumn{1}{c|}{S2020} & \multicolumn{1}{c|}{Raw}   & S2020 \\ \hline
Mazzucca (IS)                                                           & \multicolumn{1}{c|}{1.36} & \multicolumn{1}{c|}{0.13}    & \multicolumn{1}{c|}{-0.70} & 0.12    & \multicolumn{1}{c|}{-}    & \multicolumn{1}{c|}{-}       & \multicolumn{1}{c|}{-}    & -       & \multicolumn{1}{c|}{0.81} & \multicolumn{1}{c|}{-}       & \multicolumn{1}{c|}{-0.45} & -       \\ \hline
Cengio (RS)                                                             & \multicolumn{1}{c|}{0.59} & \multicolumn{1}{c|}{0.51}    & \multicolumn{1}{c|}{-0.31} & 0.26    & \multicolumn{1}{c|}{0.45} & \multicolumn{1}{c|}{0.41}    & \multicolumn{1}{c|}{0.48} & 0.15    & \multicolumn{1}{c|}{0.48} & \multicolumn{1}{c|}{0.56}    & \multicolumn{1}{c|}{-0.09} & 0.35    \\ \hline
Maggiolina (US)                                                         & \multicolumn{1}{c|}{0.69} & \multicolumn{1}{c|}{0.81}    & \multicolumn{1}{c|}{-0.43} & 0.33    & \multicolumn{1}{c|}{0.37} & \multicolumn{1}{c|}{0.51}    & \multicolumn{1}{c|}{0.36} & 0.10    & \multicolumn{1}{c|}{0.47} & \multicolumn{1}{c|}{0.62}    & \multicolumn{1}{c|}{-0.30} & 0.32    \\ \hline
Quarto (US)                                                             & \multicolumn{1}{c|}{0.53} & \multicolumn{1}{c|}{0.83}    & \multicolumn{1}{c|}{0.20}  & 0.43    & \multicolumn{1}{c|}{0.28} & \multicolumn{1}{c|}{0.34}    & \multicolumn{1}{c|}{0.16} & 0.07    & \multicolumn{1}{c|}{-}    & \multicolumn{1}{c|}{0.47}    & \multicolumn{1}{c|}{-}     & 0.31    \\ \hline
Varaldo (US)                                                            & \multicolumn{1}{c|}{-}    & \multicolumn{1}{c|}{-}       & \multicolumn{1}{c|}{-}     & -       & \multicolumn{1}{c|}{0.15} & \multicolumn{1}{c|}{0.47}    & \multicolumn{1}{c|}{0.05} & 0.08    & \multicolumn{1}{c|}{0.61} & \multicolumn{1}{c|}{0.75}    & \multicolumn{1}{c|}{-0.22} & 0.34    \\ \hline
\end{tabular}}
\label{tab:indices-selec}
\end{table}

To this aim,  Tab.\ \ref{tab:indices-selec} reports the statistical error indices for a sub-set of five control stations, classified as industrial (IS), rural (RS) and urban (US) stations. We focused on such a sub-set so to have the possibility to compare our results with
those obtained by \cite{sartini2020emission} using the high-resolution numerical strategy, called LINEA, implemented and managed by ARPA Liguria to forecast the concentration of photochemical pollutants.
In the table only daily means are considered as done  in \cite{sartini2020emission}. The added value brought by the concomitant action of high-resolution and more detailed emission information is not clearly evident. Although a detailed comparison among the two strategies is out of the scope of the present paper, we can say that neither of the two strategies looks significantly better than  the other.\\
The question addressed here is the following: can a calibration strategy properly set for
ensemble prediction systems substantially increase the forecast skill thereby overcoming persistence?\\
Answer this question is the main aim of the next sections of the present paper. As a  further aim
we will also explore the possibility to obtain calibrated forecasts significantly more accurate than those one can obtain with costly high-resolution models which also account detailed emission information made available at regional scale. Both answers have clear impacts on the selection of optimal environmental monitoring strategies where the
optimality has also  to account for the economic cost of the strategy. 

\section{A dynamic calibration strategy for multi-model ensemble}
\label{sec:dynamic}
The calibration strategy used in the present activity has its roots in the well-established
Ensemble Model Output Statistics (EMOS) through which a raw ensemble forecast can be accurately transformed into a predictive probability density function, with a simultaneous  correction of biases and dispersion errors \citep{gneiting2005calibrated,thorarinsdottir2010probabilistic}.
If on the one hand  the technique is widely used to calibrate ensemble meteorological models with remarkable improvements with respect to raw forecasts, the same technique is still in its infancy in the realm of air quality modelling. The reason is probably due to the fact that if on the side of meteorological models several high-quality ensemble forecast systems  are operative in many leading weather centres, the same situation does not apply for operative ensemble systems focused on air quality forecasts.\\
The technique we exploit here is largely inspired by a recent work of ours where it was presented and applied to calibrate wind speed at the SYNOP stations covering the Italian territory, with excellent results \citep{casciaro2022increasing}. The strategy is intimately dynamics in that it uses real time/quasi real time observed data at station to reduce the real-time  model error and generalizes the static approach proposed in a previous paper by \cite{casciaro2022novel} where observed data only enter in a training set  of past forecast-observation pair at stations. \\
To summarize the core of our strategy, called EMOS$_{+4r}$, let us consider $X_i^j$ the i-th ensemble member forecast of K members on the j-th model grid-point ($j = 1,\cdots, 4$ spans over the nearest model grid-points to the ground station) and $S^{2\:j}$ the members variance on the j-th model grid-point. We also denote by $Z_1, \cdots, Z_q$ the $q$ categorical variables (here, the 2-m temperature,  the persistence built from the last available pollutant observation, the CAMS ensemble mean for the predictant pollutant, the hour of the day, the surface wind gust, and the 10-m wind speed)
expected to be useful to disentangle the forecast error.
Meteorological conditioning variables are obtained here from the WRF numerical weather prediction
model \citep{skamarock2005description} we have run for the years 2019 and 2020 in the model 
configuration detailed in \cite{cassola2015numerical}.\\
The EMOS$_{+4r}$ strategy defines a predictive probability distribution, a gamma distribution in our strategy, with mean and variance given by
\begin{equation}
  \mu   = a(Z_1, \cdots, Z_q) + \sum_{i=1,j=1}^{M,4} b_{ij}(Z_1, \cdots, Z_q) X_i^j 
\label{eq:EMOS_mu_cond}
\end{equation}
\begin{equation}
  \sigma^2   = c(Z_1, \cdots, Z_q) + \sum_{j=1}^{4} d_{j}(Z_1, \cdots, Z_q) S^{2 \: j} 
\label{eq:EMOS_sigma_cond}
\end{equation}
where $j$ spans  over  the  4  model  grid  points  around  the  station.\\
The free parameters $a$, $b$, $c$, and $d$, actually non-parametric functions of the conditioning variables,  must be best-fitted for each combination of classes’ levels, via a training set  (the year 2019 in the present study), by minimizing the `distance' between observed data and the associated probability of occurrence. In plain terms, the quantity to be minimized is the so-called
Continuous Ranked Probability Score (CRPS) \citep{hersbach2000decomposition,gneiting2005calibrated}.\\
For a gamma distribution, a closed form for the CRPS  has been obtained by
\cite{scheuerer2015probabilistic} making the minimization procedure easy and fast. For an observation-forecast pair ($Y, \mathbf{X}$) it reads:
\begin{equation}
crps =  Y\left[2P\left(k, \frac{Y}{\theta}\right)-1\right]-k \theta\left[2P\left(k+1, \frac{Y}{\theta}\right)-1\right]-\frac{\theta}{\beta\left(\frac{1}{2},k\right)}
\label{eq:crps}
\end{equation}
with $Y$ being the observation, $P$ the incomplete gamma function \citep{abramowitz1948handbook}, and $\beta$ the beta function. The forecast vector $\mathbf{X} = (X_1, \cdots, X_M)$ comes into the expression (\ref{eq:crps}) via the parameters $k$ and $\theta$.
The quantity to be minimized in a training set where both observations and forecasts are available is:
\begin{equation}
CRPS = \frac{1}{N}\sum_{i = 1}^N crps(\mathbf{X}_i, Y_i)
\label{eq:CRPS_tot}
\end{equation}
with $i$ denoting the i-th observation-forecast pair and $N$  is the total number of pairs in the training set.\\
The CRPS combines calibration and informativeness in one index, thus 
allowing  the evaluation of predictive performance 
that is based on the paradigm of maximizing the sharpness of the predictive distributions subject to calibration \citep{gneiting2007probabilistic}.\\
The EMOS$_{+4r}$ static calibration is completed by
applying downstream to the aforementioned steps a further EMOS
with a 40-day rolling training set without conditioning on $\mathbf{Z}$.\\
How to add a dynamic character to the present strategy is described in detail by 
\cite{casciaro2022increasing}. Here we only remember that the basic idea of the strategy
is to insert two different persistences  as new additional predictors in (\ref{eq:EMOS_mu_cond}) and (\ref{eq:EMOS_sigma_cond}).  
One persistence is built from station observation at hour 9 UTC (the choice for this specific hour
can be modified as desired) while the second type of persistence accounts for the possible
occurrence of diurnal cycles. For a leading time h, it is thus defined as the observed data at the
hour h-24. Using persistences as new predictors allows one to synchronize forecasts to the nearest
observed data with benefits visible in the first few forecast hours as we will detail in the following. \\
We call our dynamic calibration D-EMOS$_{+4r}$ which exactly follows the steps reported by \cite{casciaro2022increasing}.


\section{Results}
\label{Sec:results}
\subsection{Choice of conditioning variables}
\label{sec:condi}
The following set of conditioning variables has been considered in our study for the predictands O$_3$, NO$_2$ and CO: 
the 2-m temperature (T$_{2m}$), the persistence using the predictand  observations at 9 UTC (Pers),
the mean of the CAMS predictand forecasts (Ens mean), the hour of the day (h), the surface wind gust (wg), and the 10-m wind speed (ws10).\\
For each station, the most effective three conditioning variables (i.e.\ M=3) for which the minimum CRPS is obtained are determined via a cross validation. This latter has been carried out on the training set corresponding to the year 2019, with the year 2020 serving as test set.
\begin{figure}[h!]
\includegraphics[width=\textwidth]{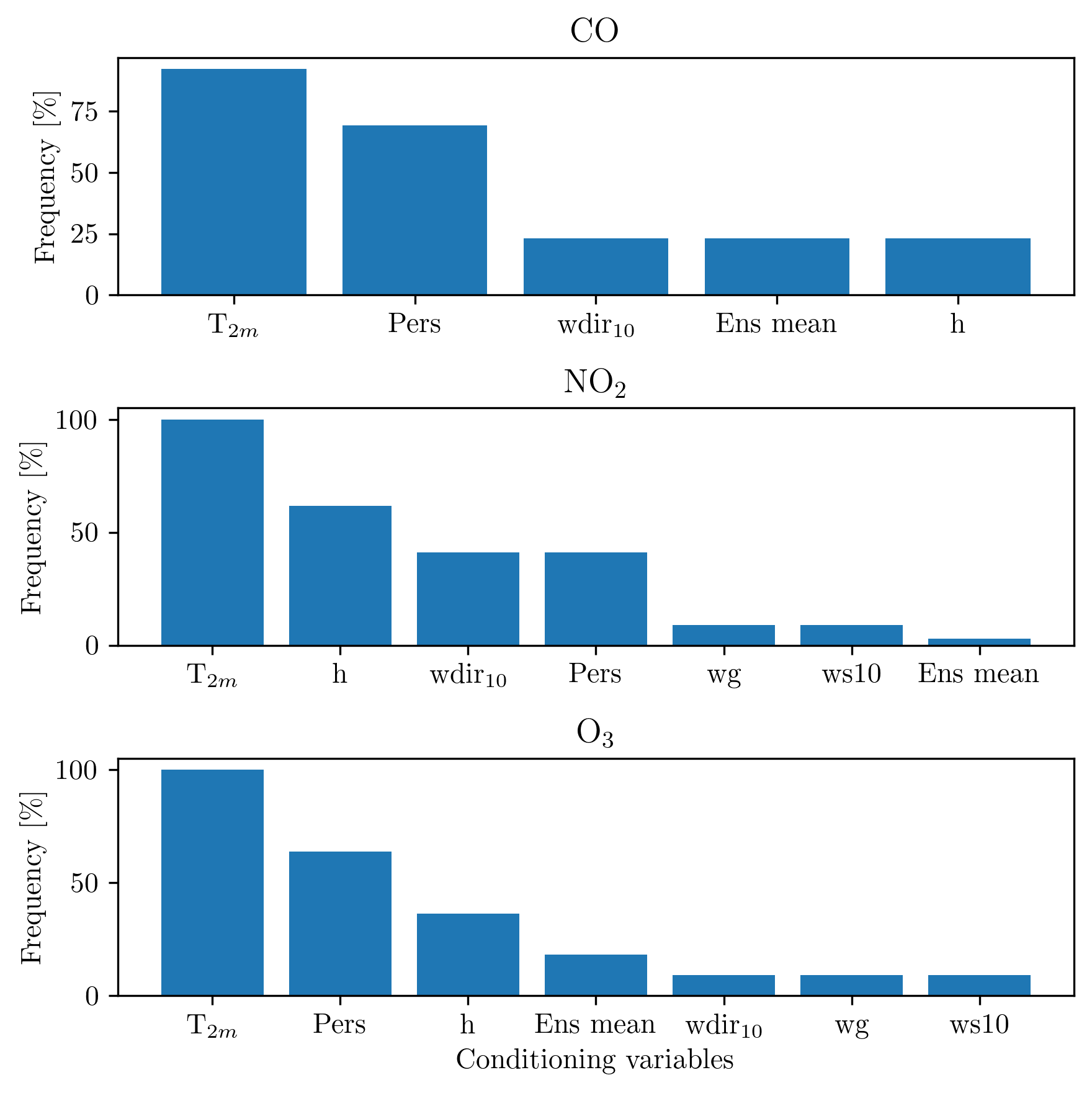}
\centering
\caption{Shown is the frequency of occurrence of the conditioning variables for CO, NO$_2$ and O$_3$. The conditioning variables we have considered  are:
the 2-m temperature (T$_{2m}$), the persistence using the predictand  observation at 9 UTC (Pers),
the mean of the CAMS predictand forecasts (Ens mean), the hour of the day (h), the surface wind gust (wg), and the 10-m wind speed (ws10).}
\label{fig:condizionamenti_selezionati}
\end{figure}
Figure \ref{fig:condizionamenti_selezionati} shows the frequency with which each conditioning variable is selected. A given percentage means the number of stations where a certain conditioning variable has been selected divided by the number of stations considered for the specific pollutant. From the figure it appears that the most important conditioning variable is the 2-m temperature signaling
a role of the thermal structure of the atmospheric boundary layer in disentangling the forecast error.
The same seems to be confirmed by the important role of the hour of the day as conditioning variable.\\
For the PMs, the importance of conditioning variables turned out to be negligible. For this reason,
the calibration strategy for the PMs will proceed by assuming M=0 in the algorithm.

\subsection{Assessing the added value of our calibration}
\label{sec:assess}
\begin{figure}[h!]
\includegraphics[width=\textwidth]{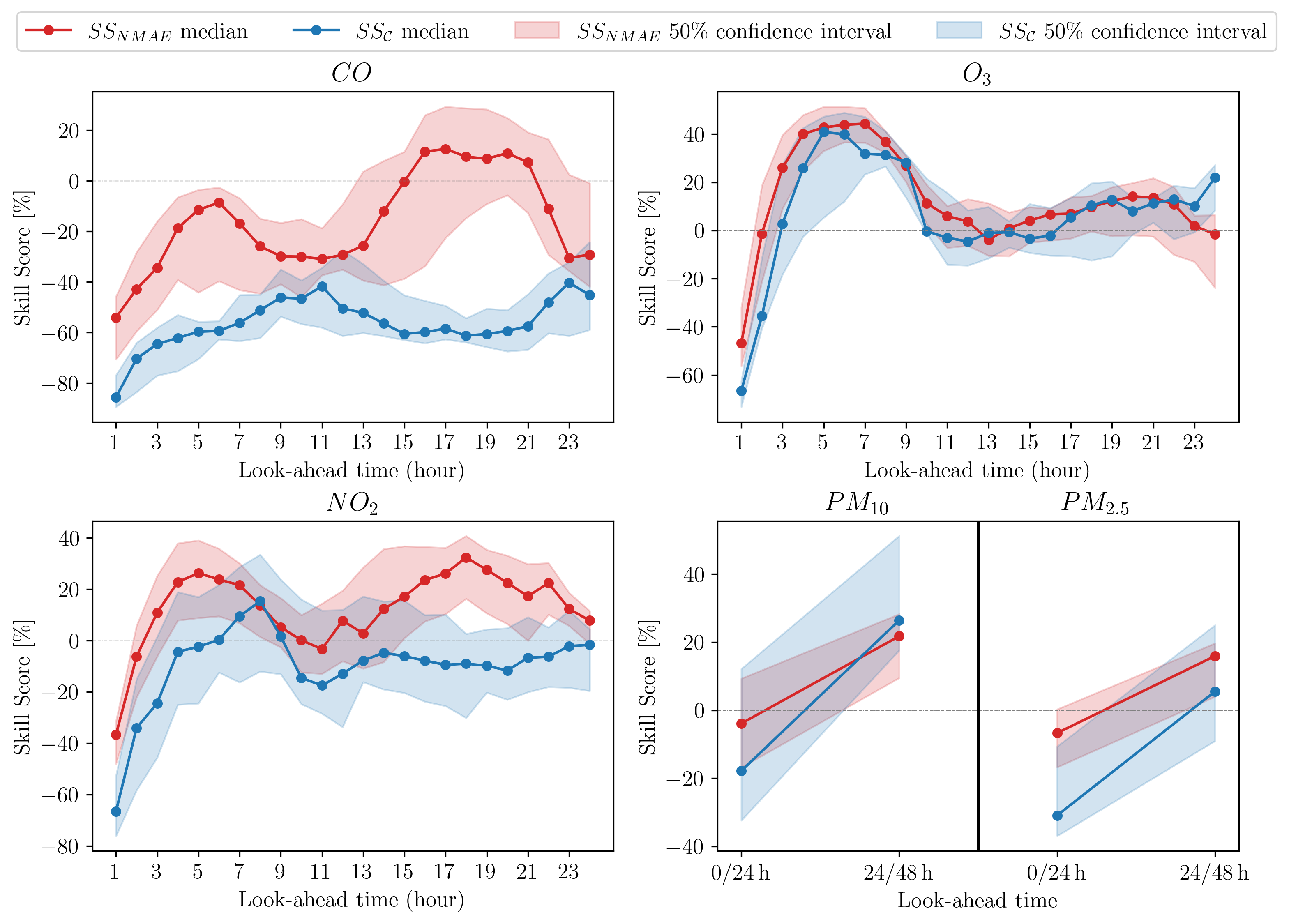}
\centering
\caption{As in Fig. \ref{fig:raw_vs_pers00} but now the skill of the EMOS$_0$-based calibration is assessed against persistence built from the 9 UTC observation.}
\label{fig:E0_vs_pers00}
\end{figure}

\begin{figure}[h!]
\includegraphics[width=\textwidth]{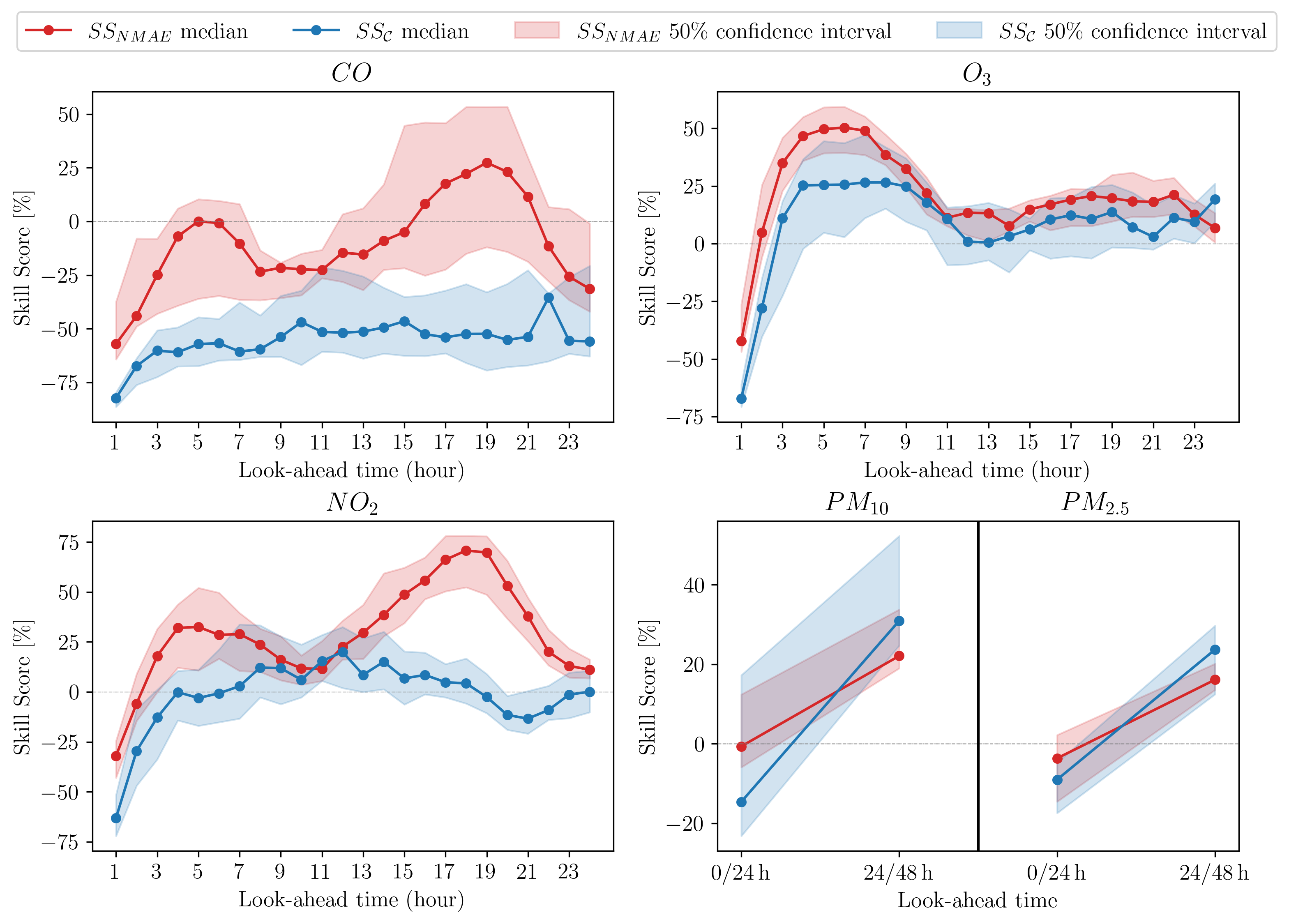}
\centering
\caption{
As in Fig. \ref{fig:raw_vs_pers00} but now the skill of our static calibration EMOS$_{+4r}$ is assessed against persistence.}
\label{fig:stat_vs_pers00}
\end{figure}

\begin{figure}[h!]
\includegraphics[width=\textwidth]{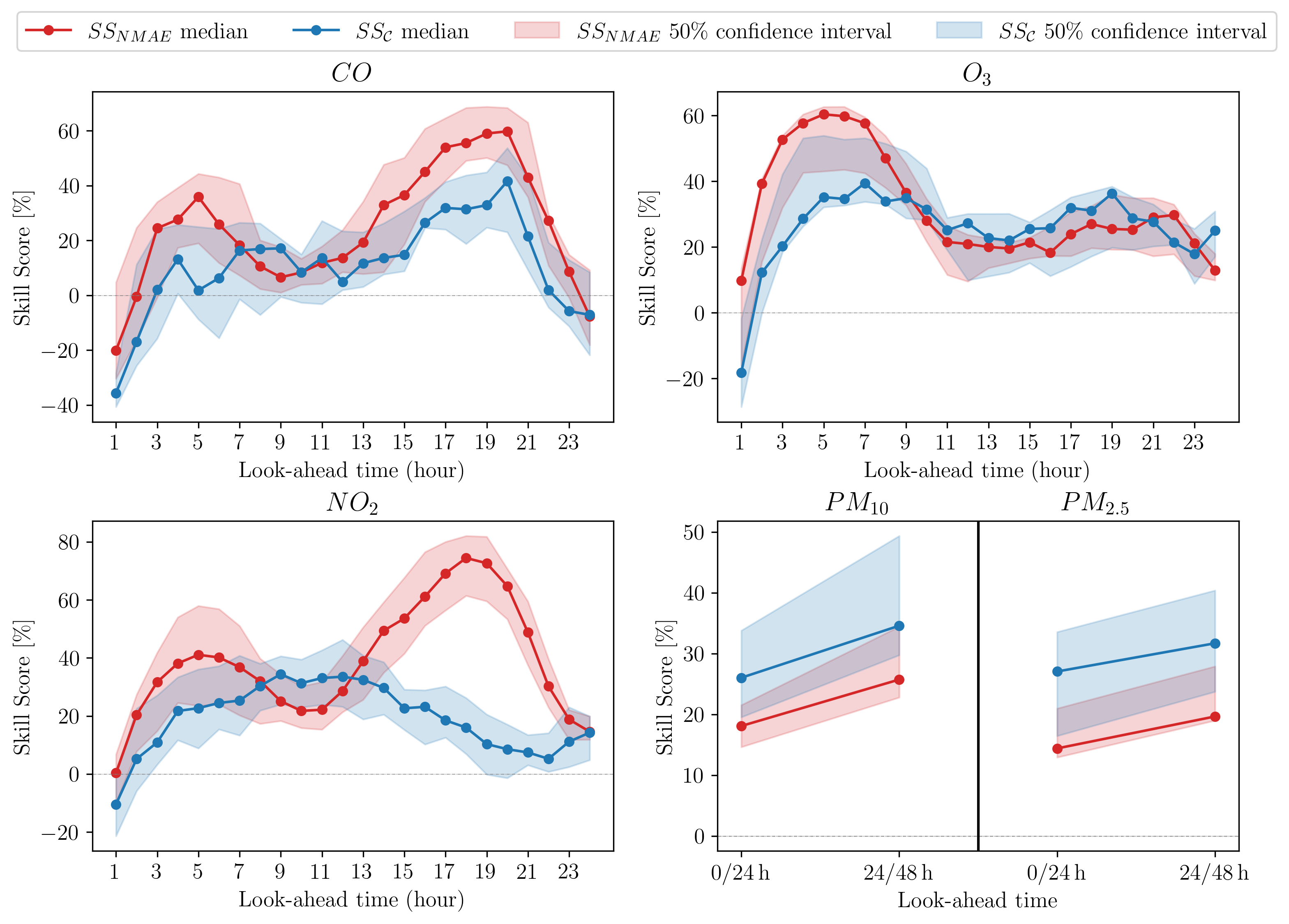}
\centering
\caption{
As in Fig. \ref{fig:raw_vs_pers00} but now the skill of our dynamic calibration D-EMOS$_{+4r}$ is assessed against 9 UTC persistence.}
\label{fig:dyn_vs_pers00}
\end{figure}

\begin{figure}[h!]
\includegraphics[width=\textwidth]{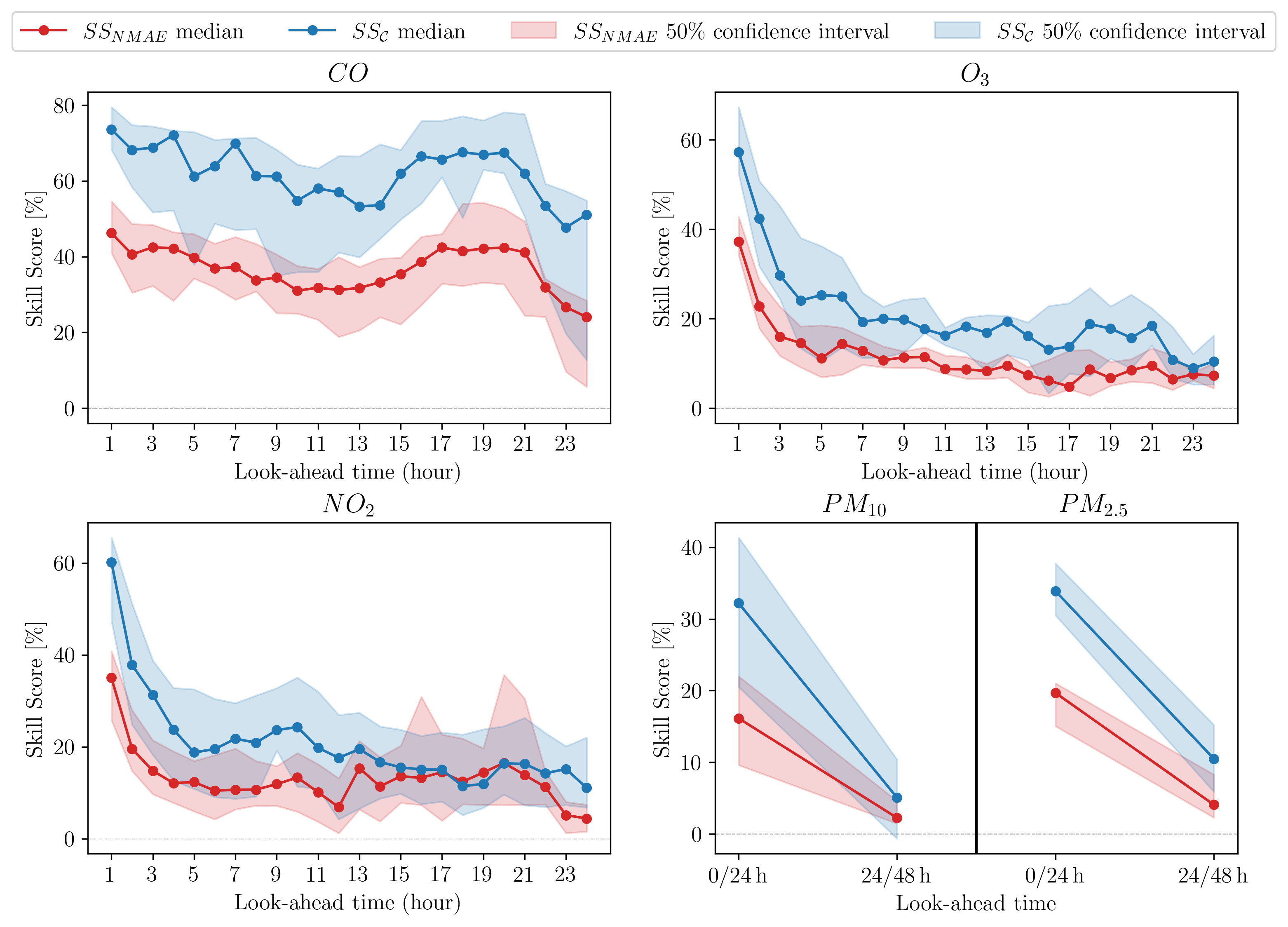}
\centering
\caption{
As in Fig. \ref{fig:raw_vs_pers00} but now the skill of our dynamic calibration D-EMOS$_{+4r}$ is assessed against static calibration EMOS$_{+4r}$.}   
\label{fig:dyn_vs_stat}
\end{figure}

\begin{figure}[h!]
\includegraphics[width=\textwidth]{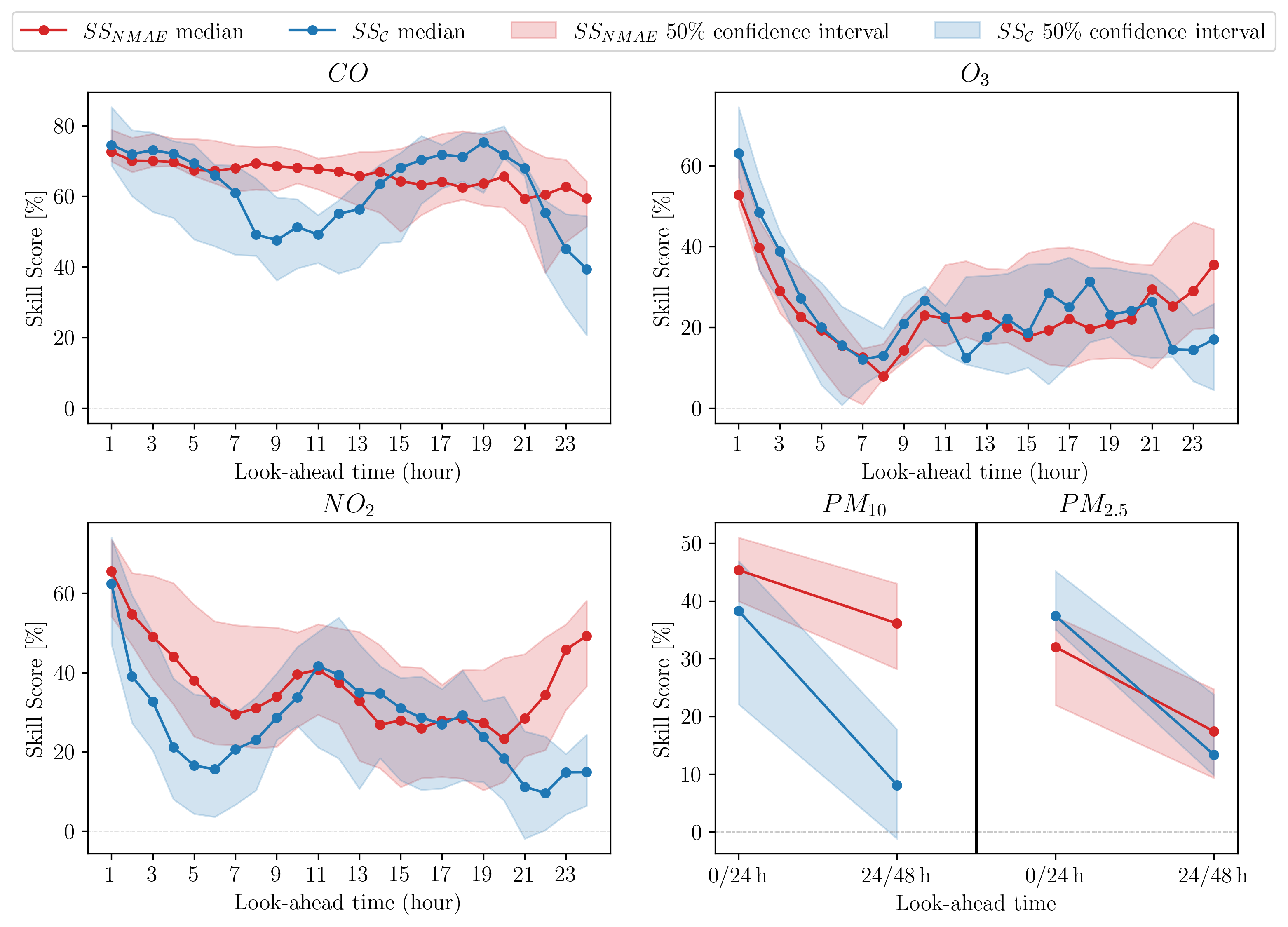}
\centering
\caption{
As in Fig. \ref{fig:raw_vs_pers00} but now the skill of our dynamic calibration D-EMOS$_{+4r}$ is assessed against raw forecast (mean of CAMS members).}
\label{fig:dyn_vs_raw}
\end{figure}

\begin{figure}[h!]
\includegraphics[width=\textwidth]{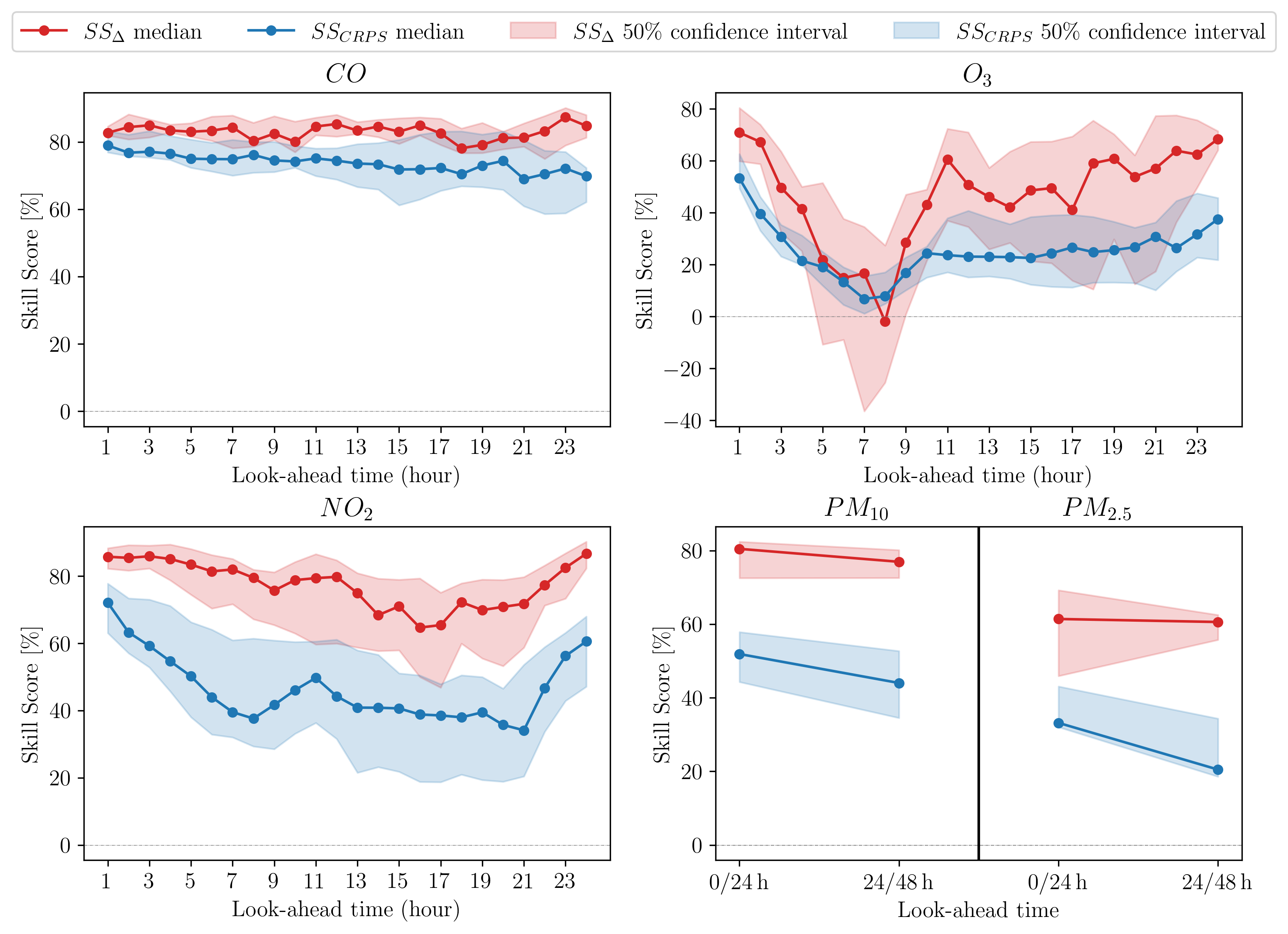}
\centering
\caption{
As in Fig. \ref{fig:raw_vs_pers00} but now the skill of D-EMOS$_{+4r}$ is assessed in terms of indices $\Delta$ and CRPS using the raw forecasts (CAMS members mean) as reference.}
\label{fig:dyn_vs_raw_prob}
\end{figure}

Having selected the most effective three conditioning variables for O$_3$, NO$_2$ and CO, we are now ready to assess the skill of our calibrated forecasts  using the year 2020 as test set. Let us start by considering the simplest static EMOS strategy where neither the conditioning variables  nor the rolling training set are considered, and the forecast is for the model grid point closest to the station.  Let us call this baseline calibration EMOS$_0$. The skill of the EMOS$_0$-based calibration is shown in Fig.\ \ref{fig:E0_vs_pers00} using 9 UTC persistence as a reference.
Although improvements with respect to the raw forecasts shown in
Fig.\ \ref{fig:raw_vs_pers00} are clearly detectable, persistence is still the best
forecast for CO, for almost all look-ahead times, and for the
shortest look-ahead  times for all other pollutants.\\
These remarks suggest the use of more elaborated calibrations. Let us start
from our static EMOS$_{+4r}$ calibration the added value of which, against persistence, is shown in Fig.\ \ref{fig:stat_vs_pers00}.
A further improvement is clearly observed downstream of the  EMOS$_{+4r}$ calibration even if problems
still remain. In particular, for CO persistence remains the best choice for all forecast horizons even if it is losing its lead. In all cases persistence continues to overcome the calibrated forecasts
for the shortest forecast horizons. \\
This last remark strongly suggests the use of real-time data to correct the model error in its
infancy. This is done here in terms of our dynamic calibration D-EMOS$_{+4r}$ the skill of which is shown in Fig.\ \ref{fig:dyn_vs_pers00} using 9 UTC persistence as a reference.
A net change of paradigm brought by our dynamic calibration D-EMOS$_{+4r}$ is detectable.
Persistence is no longer the best choice for CO and the calibrated forecast increases its lead over persistence for all pollutants considered.\\
To better isolate the added value of our dynamic calibration D-EMOS$_{+4r}$ against our static EMOS$_{+4r}$ calibration, in Fig.\ \ref{fig:dyn_vs_stat}
we have reported the skill of the  D-EMOS$_{+4r}$-based forecasts using
the EMOS$_{+4r}$-based forecasts as a reference.
The improvement of D-EMOS$_{+4r}$ relative to EMOS$_{+4r}$ is evident for all look-ahead horizons, being more pronounced, as expected, for the shortest forecast horizons.\\
Our best calibration D-EMOS$_{+4r}$ can be finally assessed with respect to the raw forecasts in
Fig.\ \ref{fig:dyn_vs_raw}.
The enhanced quality of the calibrated forecasts with respect to the raw forecasts is
noteworthy. A similar conclusion holds for the assessment of calibration and informativeness
encoded in the indices CRPS and $\Delta$ (see \ref{App:stat}), as one can see
from Fig.\ \ref{fig:dyn_vs_raw_prob}.

From all figures shown in this section we can answer the question we addressed in Sec.\ \ref{sec:EPS} on whether a calibration strategy properly set for
ensemble prediction systems can substantially increase the skill of air quality multi-model forecasts. The answer is definitely positive and a key role played by real-time observations clearly emerge.

We are now ready to answer the last question addressed in Sec.\ \ref{sec:EPS}. It can be rephrased
here as follow. Is our best calibration competitive, or even better, when compared against
forecasts produced by raw
high-resolution air quality models which also 
account for detailed emission information made available at regional scale?\\

\begin{table}[h!]
\centering
\caption{As in Tab.\ \ref{tab:indices-selec} but now the mean of the raw CAMS members is replaced by our dynamic calibration D-EMOS$_{+4r}$ (D-EMOS in short).}
\resizebox{\textwidth}{!}{%
\begin{tabular}{|c|cccc|cccc|cccc|}
\hline
\multirow{3}{*}{\begin{tabular}[c]{@{}c@{}}Station\\ name\end{tabular}} & \multicolumn{4}{c|}{NO2}                                                                           & \multicolumn{4}{c|}{O3}                                                                            & \multicolumn{4}{c|}{PM10}                                                                          \\ \cline{2-13} 
                                                                        & \multicolumn{2}{c|}{HH}                                    & \multicolumn{2}{c|}{NBI}              & \multicolumn{2}{c|}{HH}                                    & \multicolumn{2}{c|}{NBI}              & \multicolumn{2}{c|}{HH}                                    & \multicolumn{2}{c|}{NBI}              \\ \cline{2-13} 
                                                                        & \multicolumn{1}{c|}{D-EMOS} & \multicolumn{1}{c|}{S2020} & \multicolumn{1}{c|}{D-EMOS} & S2020 & \multicolumn{1}{c|}{D-EMOS} & \multicolumn{1}{c|}{S2020} & \multicolumn{1}{c|}{D-EMOS} & S2020 & \multicolumn{1}{c|}{D-EMOS} & \multicolumn{1}{c|}{S2020} & \multicolumn{1}{c|}{D-EMOS} & S2020 \\ \hline
Mazzucca (IS)                                                           & \multicolumn{1}{c|}{0.25}   & \multicolumn{1}{c|}{0.13}    & \multicolumn{1}{c|}{0.04}   & 0.12    & \multicolumn{1}{c|}{-}      & \multicolumn{1}{c|}{-}       & \multicolumn{1}{c|}{-}      & -       & \multicolumn{1}{c|}{0.28}   & \multicolumn{1}{c|}{-}       & \multicolumn{1}{c|}{0.02}   & -       \\ \hline
Cengio (RS)                                                             & \multicolumn{1}{c|}{0.32}   & \multicolumn{1}{c|}{0.51}    & \multicolumn{1}{c|}{0.00}   & 0.26    & \multicolumn{1}{c|}{0.18}   & \multicolumn{1}{c|}{0.41}    & \multicolumn{1}{c|}{0.02}   & 0.15    & \multicolumn{1}{c|}{0.39}   & \multicolumn{1}{c|}{0.56}    & \multicolumn{1}{c|}{-0.01}  & 0.35    \\ \hline
Maggiolina (US)                                                         & \multicolumn{1}{c|}{0.23}   & \multicolumn{1}{c|}{0.81}    & \multicolumn{1}{c|}{0.03}   & 0.33    & \multicolumn{1}{c|}{0.17}   & \multicolumn{1}{c|}{0.51}    & \multicolumn{1}{c|}{0.04}   & 0.10    & \multicolumn{1}{c|}{0.26}   & \multicolumn{1}{c|}{0.62}    & \multicolumn{1}{c|}{0.03}   & 0.32    \\ \hline
Quarto (US)                                                             & \multicolumn{1}{c|}{0.39}   & \multicolumn{1}{c|}{0.83}    & \multicolumn{1}{c|}{0.09}   & 0.43    & \multicolumn{1}{c|}{0.17}   & \multicolumn{1}{c|}{0.34}    & \multicolumn{1}{c|}{0.03}   & 0.07    & \multicolumn{1}{c|}{-}      & \multicolumn{1}{c|}{0.47}    & \multicolumn{1}{c|}{-}      & 0.31    \\ \hline
Varaldo (US)                                                            & \multicolumn{1}{c|}{-}      & \multicolumn{1}{c|}{-}       & \multicolumn{1}{c|}{-}      & -       & \multicolumn{1}{c|}{0.13}   & \multicolumn{1}{c|}{0.47}    & \multicolumn{1}{c|}{-0.01}  & 0.08    & \multicolumn{1}{c|}{0.35}   & \multicolumn{1}{c|}{0.75}    & \multicolumn{1}{c|}{-0.03}  & 0.34    \\ \hline
\end{tabular}}
\label{tab-indices}
\end{table}

Table \ref{tab-indices} gives us a clear answer. Shown in the table are all error indices we have 
computed for the five control stations already considered in  Tab.\ \ref{tab:indices-selec}. The table also includes the error indices for the same stations associated to the forecast model used by \cite{sartini2020emission}.  The conclusion is that the potential benefits
brought by  the combined use of a high-resolution model and a more detailed inventory information
can be significantly exceeded  upon implementing a proper dynamic calibration of the coarser model strategy.

We conclude this section by summarizing the skill of the calibrated CAMS ensemble in terms of all considered indices. These latter have been calculated as a mean
of the indices over all stations. The results grouped together by pollutant type are reported in Tab.\ \ref{tab:tabellone_errori_medi}.


\begin{table}
  \centering
\caption{Summary of all relevant statistical error indices for the mean of the 9 CAMS members on the year 2020. Indices   have been obtained by averaging all single index over all stations. 
   Daily averages are considered for PM$_{10}$ and PM$_{2.5}$ while hourly averages for  O$_{3}$ and NO$_{2}$ and CO.}
  \begin{tabular}{|c|c|c|c|c|c|} 
\hline
          & \begin{tabular}[c]{@{}c@{}}Raw ens.\\ mean\end{tabular} & Persistence & \begin{tabular}[c]{@{}c@{}}Static\\ calibration\end{tabular} & \begin{tabular}[c]{@{}c@{}}Dynamic\\ calibration\end{tabular} &                              \\ 
\hline
NMAE      & 0.71                                                    & 0.38        & 0.38                                                         & 0.25                                                          & \multirow{6}{*}{CO}          \\ 
\cline{1-5}
$\cal{C}$ & 0.34                                                    & 0.63        & 0.41                                                         & 0.75                                                          &                              \\ 
\cline{1-5}
HH        & 1.57                                                    & 0.48        & 0.51                                                         & 0.34                                                          &                              \\ 
\cline{1-5}
NBI       & -0.69                                                   & 0.20        & -0.02                                                        & -0.02                                                         &                              \\ 
\cline{1-5}
$\Delta$  & 1.61                                                    & -           & 0.32                                                         & 0.19                                                          &                              \\ 
\cline{1-5}
CRPS      & 0.39                                                    & -           & 0.15                                                         & 0.10                                                          &                              \\ 
\hline
NMAE      & 0.27                                                    & 0.28        & 0.21                                                         & 0.18                                                          & \multirow{6}{*}{O$_3$}       \\ 
\cline{1-5}
$\cal{C}$ & 0.73                                                    & 0.62        & 0.75                                                         & 0.81                                                          &                              \\ 
\cline{1-5}
HH        & 0.30                                                    & 0.38        & 0.25                                                         & 0.22                                                          &                              \\ 
\cline{1-5}
NBI       & 0.15                                                    & -0.13       & 0.01                                                         & 0.01                                                          &                              \\ 
\cline{1-5}
$\Delta$  & 0.48                                                    & -           & 0.15                                                         & 0.14                                                          &                              \\ 
\cline{1-5}
CRPS      & 13.2                                                    & -           & 9.5                                                          & 8.4                                                           &                              \\ 
\hline
NMAE      & 0.67                                                    & 0.63        & 0.42                                                         & 0.36                                                          & \multirow{6}{*}{NO$_2$}      \\ 
\cline{1-5}
$\cal{C}$ & 0.41                                                    & 0.41        & 0.60                                                         & 0.70                                                          &                              \\ 
\cline{1-5}
HH        & 1.17                                                    & 0.69        & 0.51                                                         & 0.45                                                          &                              \\ 
\cline{1-5}
NBI       & -0.39                                                   & 0.29        & 0.04                                                         & 0.03                                                          &                              \\ 
\cline{1-5}
$\Delta$  & 0.98                                                    & -           & 0.13                                                         & 0.10                                                          &                              \\ 
\cline{1-5}
CRPS      & 11.3                                                    & -           & 5.6                                                          & 4.8                                                           &                              \\ 
\hline
NMAE      & 0.41                                                    & 0.34        & 0.34                                                         & 0.28                                                          & \multirow{6}{*}{PM$_{2.5}$}  \\ 
\cline{1-5}
$\cal{C}$ & 0.53                                                    & 0.63        & 0.55                                                         & 0.72                                                          &                              \\ 
\cline{1-5}
HH        & 0.51                                                    & 0.47        & 0.48                                                         & 0.39                                                          &                              \\ 
\cline{1-5}
NBI       & -0.14                                                   & 0.01        & 0.00                                                         & 0.01                                                          &                              \\ 
\cline{1-5}
$\Delta$  & 0.51                                                    & -           & 0.20                                                         & 0.19                                                          &                              \\ 
\cline{1-5}
CRPS      & 3.9                                                     & -           & 2.8                                                          & 2.4                                                           &                              \\ 
\hline
NMAE      & 0.42                                                    & 0.28        & 0.27                                                         & 0.22                                                          & \multirow{6}{*}{PM$_{10}$}   \\ 
\cline{1-5}
$\cal{C}$ & 0.51                                                    & 0.58        & 0.54                                                         & 0.69                                                          &                              \\ 
\cline{1-5}
HH        & 0.63                                                    & 0.39        & 0.36                                                         & 0.31                                                          &                              \\ 
\cline{1-5}
NBI       & -0.33                                                   & 0.00        & 0.02                                                         & 0.02                                                          &                              \\ 
\cline{1-5}
$\Delta$  & 0.85                                                    & -           & 0.20                                                         & 0.19                                                          &                              \\ 
\cline{1-5}
CRPS      & 6.8                                                     & -           & 3.7                                                          & 3.1                                                           &                              \\
\hline
\end{tabular}
\label{tab:tabellone_errori_medi}
\end{table}

\clearpage

\section{Conclusions and perspectives}
\label{Sec:conclu}
The CAMS air quality model has been assessed for  PM$_{10}$, PM$_{2.5}$, O$_{3}$, NO$_{2}$, and CO
against observations collected by
the Regional Monitoring Network of the Liguria region (northwestern Italy) in the years 2019 and 2020.
Liguria has a complex orography and is characterized by mutual interactions between land and sea circulations making accurate weather predictions a difficult task. The same holds true for  air quality forecasts because of the key role of weather in conditioning air quality. The selected observation network is thus a particularly selective  benchmark  to assess air quality forecast strategies.\\
The main conclusion of this first step of our analysis is that in its raw form the CAMS predictive
strength is very limited, with the forecast built from persistence having higher skills for all pollutants considered.\\
Motivated by these evidences we have asked the question on whether a suitable calibration strategy acting as a post-processing  of the raw CAMS forecasts may bring out the intrinsic strength of CAMS multi-model ensemble strategy. To answer this question we have applied a dynamic calibration strategy, named D-EMOS$_{+4r}$, from which a  positive answer to our question definitely came.  Our results also highlight the key role of pollutant real-time observations to be ingested in the calibration strategy.
They bring a clear added value especially in the shorter look-ahead forecast hours.\\
These results also point out the key value of observations, including their management to be used in real time: they serve not only to have detailed information on the current air quality
state  but also to predict with accuracy the future scenario with positive outcomes for the reliability  of pollution early warning systems. 

A second question was also answered in relation to the possible added value of higher-resolution forecast strategies ingesting detailed information from the inventory emission. When comparing the raw CAMS forecasts against the higher-resolution strategy by \cite{sartini2020emission} the aforementioned added value
hardly emerges. The situation of substantial equivalence is however totally washed out when our dynamic calibration D-EMOS$_{+4r}$ is applied to the raw CAMS forecasts, thus bringing out the intrinsic strength of CAMS multi-model ensemble strategy.
Left to future investigations, it will be interesting to see whether our strategy applied
to higher-resolution forecasts brings additional advantages.\\
Our new dynamic calibration gave birth to the operative air-quality prediction system for the Liguria region \citep{sitoweb}. 

\section{Acknowledgments}
\noindent
G.C. has been funded by the Italian bank foundation ``Fondazione Carige''. A.M. acknowledges the funding from the Interreg Italia-Francia Marittimo SICOMAR+ project (grant number D36C17000120006) and from the Compagnia di San Paolo (Project MINIERA No. I34I20000380007).
Discussions with Francesco Ferrari and Ludovica Sartini are warmly acknowledged. Observational data have been provided by ARPAL the activity of which is acknowledged.

\appendix

\section{Statistical indices}
\label{App:stat}
The following error indices are considered in the present study:
the normalized mean absolute error (NMAE),
the normalized bias (NBI), the symmetrically normalized root mean square error (HH), the correlation coefficient (${\cal C}$), and the so-called reliability index ($\Delta$) proposed by \cite{delle2006probabilistic}.
The advantage of HH with respect to the widely used non-symmetrically normalized root mean square error, RMSE, is extensively discussed by \cite{mentaschi2013problems}.

By denoting with $Y_n$ the n-th observation, $X_n$ the corresponding n-th forecast (here corresponding to the mean of the CAMS ensemble) and $N$ the number of observation-forecast pairs in a given test set (here the year 2020), the above mentioned statistical indices are defined as: 
\begin{equation}
NMAE = \frac{\sum_{n=1}^{N}\left|X_n - Y_n\right|}{\sum_{n=1}^{N}Y_n}
\label{eq:NMAE}
\end{equation}
\begin{equation}
NBI = \frac{\sum_{n=1}^{N}\left (Y_n - X_n\right) }{\sum_{n=1}^{N}Y_n}
\label{eq:NBI}
\end{equation}
\begin{equation}
HH = \left (\frac{\sum_{n=1}^{N}\left (Y_n - X_n\right)^2 }{\sum_{n=1}^{N}Y_nX_n}\right )^{1/2}
\label{eq:HH}
\end{equation}
%
\begin{equation} 
{\cal{C}} = \frac{\sum_{n=1}^{N}(X_n-\overline{X})(Y_n-\overline{Y})}{N\sigma_X \sigma_Y}
\label{eq:Pearson}
\end{equation}
with:
\begin{equation}
\sigma_X = \sqrt{\frac{\sum_{n=1}^{N}(X_n-\overline{X})^2}{N}}
\end{equation}
\begin{equation}
\sigma_Y = \sqrt{\frac{\sum_{n=1}^{N}(Y_n-\overline{Y})^2}{N}}
\end{equation}
where $\overline{X}$ and $\overline{Y}$ are the mean values of $X$ and $Y$.
\\
The goal of the probabilistic forecast, according to \cite{gneiting2007probabilistic}, is to maximize the sharpness of the predictive distribution subject to calibration. \cite{anderson1996method} and \cite{hamill1997verification} proposed the use of verification rank (VR) histograms to assess the calibration of ensemble forecasts. VR histograms show the distribution of the ranks when the ranks of the observations are pooled within the ordered ensemble forecasts.
In a calibrated ensemble, the observations and ensemble predictions should be interchangeable, resulting in a uniform VR histogram. The continuous analogue of the VR histogram is the probability integral transform (PIT) histogram \citep{dawid1984present, diebold1997evaluating, gneiting2007probabilistic}. The PIT value is determined by the value of the predictive cumulative distribution function at the verifying observation. For calibrated forecasts, the empirical cumulative distribution function of PIT values should converge to the uniform distribution.\\
The reliability index $\Delta$ was proposed by \cite{delle2006probabilistic} to quantify the deviation of VR histograms from uniformity. To quantify the deviation from uniformity in the PIT histograms, we use here the following definition of $\Delta$:

\begin{equation} 
\Delta = \sum_{i=1}^{m}\left|f_i - \frac{1}{m}\right|
\label{eq:Delta}
\end{equation}
where $m$ is the histogram  number of classes, each with a relative frequency of $1/m$, and $f_i$ is the observed relative frequency in class $i$.\\
This index ranges from 0 to $+\infty$, with the bound zero corresponding to
optimal forecast.\\


Having defined the statistical error indices,
a symmetric variant of the standard Skill Score  index \citep{wilks2011statistical} is used here to make the comparison between different forecasts (raw or calibrated) as quantitative as possible. Our new definition  is used to assess the performance 
of a given forecast by comparing its associated statistical error index against the one corresponding to a reference forecast. More quantitatively,
the symmetrized skill score, SS, is defined here as 
\[ 
SS = \frac{A - A_{ref}}{A_{opt} - A_{ref}}\qquad  \mbox{when}\qquad  \frac{A - A_{ref}}{A_{opt} - A_{ref}} \ge 0
\]
\[
SS = -\frac{A_{ref}-A}{A_{opt} - A}\qquad   \mbox{elsewhere}
\]
where $A$ is the value of a suitable error index associated to forecast, $A_{ref}$ is the same as $A$ but relative to a reference forecast. Finally, $A_{opt}$ refers to the optimal
index value here supposed to be 0 (e.g.\ as for NMAE) or 1 (e.g.\ as for the correlation coefficient). All these error indices are supposed here to be positive (e.g.\ as it is for the NMAE and for the correlation coefficient apart situations of anticorrelated regimes however not of interest here).

A perfect calibration yields SS=1, corresponding to the upper bound of SS. Values of SS smaller than one (including negative values) indicate that the forecast is less accurate than the reference one.

\section{Assessment of the skill of the single CAMS members for the year 2020}
\label{Confronto modelli raw} 
For the sake of comparison, we assess here the skill of the 9 CAMS ensemble members.
Error indices  have been calculated for the year 2020 for 
PM$_{10}$, PM$_{2.5}$ (daily averages), O$_{3}$, NO$_{2}$, and CO (hourly averages) against observations
from the Ligurian monitoring network.
CAMS median and mean are also assessed together with the 9 members.
While for the majority of indices it is hard to identify a model clearly overcoming the others, for the correlation coefficient CAMS mean and median show the highest skill, thus confirming the power of the ensemble approach. 

\begin{table}[h!]
\centering
\caption{Statistical error indices for the 9 raw CAMS members: CHIMERE (CH) from INERIS (France), EMEP (EM) from MET Norway (Norway), EURAD-IM (EU) from J\"ulich IEK (Germany), LOTOS-EUROS (LO) from KNMI and TNO (Netherlands), MATCH (MA) from SMHI (Sweden), MOCAGE (MO) from Meteo-France (France), SILAM (S) from FMI (Finland), DEHM (DE) from Aarhus University (Denmark), and GEM-AQ (GE) from IEP-NRI (Poland). The ensemble mean and median are, Mean and Median, respectively.}
\resizebox{\textwidth}{!}{%
\begin{tabular}{|c|c|c|c|c|c|c|c|c|c|c|c|c|}
\hline
     & EM    & SI    & MO    & EU    & LO    & MA    & CH    & GE    & DE    & Median    & Mean    &                        \\ \hline
NMAE & 0.69  & 0.68  & 0.70  & 0.73  & 0.71  & 0.73  & 0.71  & 0.69  & 0.73  & 0.71  & 0.71  & \multirow{4}{*}{CO}    \\ \cline{1-12}
HH   & 1.5   & 1.4   & 1.6   & 1.7   & 1.6   & 1.7   & 1.6   & 1.5   & 1.7   & 1.6   & 1.6   &                        \\ \cline{1-12}
NBI  & -0.66 & -0.66 & -0.68 & -0.72 & -0.69 & -0.72 & -0.69 & -0.66 & -0.71 & -0.70 & -0.69 &                        \\ \cline{1-12}
$\cal{C}$    & 0.17  & 0.31  & 0.29  & 0.20  & 0.24  & 0.19  & 0.22  & 0.21  & 0.23  & 0.26  & 0.27  &                        \\ \hline
NMAE & 0.29  & 0.31  & 0.41  & 0.36  & 0.35  & 0.36  & 0.38  & 0.37  & 0.36  & 0.30  & 0.29  & \multirow{4}{*}{O$_3$}    \\ \cline{1-12}
HH   & 0.33  & 0.39  & 0.42  & 0.42  & 0.37  & 0.37  & 0.39  & 0.41  & 0.38  & 0.32  & 0.31  &                        \\ \cline{1-12}
NBI  & 0.12  & -0.09 & 0.31  & 0.11  & 0.22  & 0.28  & 0.28  & 0.21  & 0.25  & 0.20  & 0.19  &                        \\ \cline{1-12}
$\cal{C}$    & 0.66  & 0.63  & 0.58  & 0.41  & 0.62  & 0.65  & 0.60  & 0.58  & 0.57  & 0.71  & 0.73  &                        \\ \hline
NMAE & 0.68  & 0.69  & 0.78  & 0.71  & 0.73  & 0.74  & 0.76  & 0.75  & -     & 0.68  & 0.67  & \multirow{4}{*}{NO$_2$}   \\ \cline{1-12}
HH   & 1.3   & 0.99  & 1.1   & 1.4   & 1.3   & 1.7   & 1.3   & 1.3   & -     & 1.3   & 1.2   &                        \\ \cline{1-12}
NBI  & -0.43 & -0.17 & -0.13 & -0.47 & -0.43 & -0.63 & -0.34 & -0.34 & -     & -0.43 & -0.37 &                        \\ \cline{1-12}
$\cal{C}$    & 0.33  & 0.38  & 0.38  & 0.31  & 0.28  & 0.30  & 0.16  & 0.23  & -     & 0.37  & 0.38  &                        \\ \hline
NMAE & 0.58  & 0.52  & 0.47  & 0.48  & 0.55  & 0.44  & 0.46  & 0.57  & 0.46  & 0.41  & 0.40  & \multirow{4}{*}{PM$_{2.5}$} \\ \cline{1-12}
HH   & 0.72  & 0.64  & 0.84  & 0.67  & 0.93  & 0.69  & 0.58  & 0.72  & 0.62  & 0.61  & 0.57  &                        \\ \cline{1-12}
NBI  & 0.11  & 0.16  & -0.39 & -0.07 & -0.41 & -0.24 & 0.05  & 0.21  & -0.11 & -0.13 & -0.08 &                        \\ \cline{1-12}
$\cal{C}$    & 0.43  & 0.50  & 0.51  & 0.45  & 0.40  & 0.50  & 0.49  & 0.38  & 0.55  & 0.53  & 0.55  &                        \\ \hline
NMAE & 0.46  & 0.39  & 0.61  & 0.48  & 0.54  & 0.48  & 0.44  & 0.48  & 0.42  & 0.44  & 0.41  & \multirow{4}{*}{PM$_{10}$}  \\ \cline{1-12}
HH   & 0.65  & 0.55  & 1.14  & 0.77  & 0.83  & 0.75  & 0.65  & 0.74  & 0.61  & 0.59  & 0.63  &                        \\ \cline{1-12}
NBI  & -0.22 & -0.17 & -0.60 & -0.41 & -0.32 & -0.41 & -0.29 & -0.29 & -0.31 & -0.37 & -0.34 &                        \\ \cline{1-12}
$\cal{C}$    & 0.44  & 0.52  & 0.49  & 0.47  & 0.41  & 0.50  & 0.43  & 0.32  & 0.56  & 0.54  & 0.55  &                        \\ \hline
\end{tabular}}
\label{tab:tabellone_errori_medi_all}
\end{table}



\bibliographystyle{elsarticle-harv} 
\bibliography{biblio.bib}

\end{document}